\documentclass[10pt]{article}
 
\usepackage{fullpage} 
\usepackage{amsfonts}
\usepackage[]{algorithm}
\usepackage{algorithmic}
\usepackage{graphicx}
\usepackage{verbatim}
\usepackage{longtable}
\usepackage{supertabular}
\usepackage{url}
\usepackage{hyperref}

\graphicspath{{Fig/}}

\def\HC{\mathrm{HC}}
\def\calM{\mathcal{M}}

\begin{document}
 
\title{Clustering patterns connecting COVID-19 dynamics and Human mobility using optimal transport }
 
\author{Frank~Nielsen\thanks{F. Nielsen is with Sony Computer Science Laboratories Inc, Tokyo, Japan. E-mail: {\tt Frank.Nielsen@acm.org}}
\and  Gautier~Marti\thanks{G. Marti is an independent researcher. E-mail: {\tt gautier.marti@gmail.com}}
\and Sumanta~Ray\thanks{S. Ray is with (1) Centrum Wiskunde \& Informatica, Science Park 123, 1098 XG Amsterdam, The Netherlands. (2) Department of Computer Science and Engineering, Aliah University, Kolkata, India. e-mail: {\tt Sumanta.Ray@cwi.nl}}
\and  Saumyadipta Pyne 
\thanks{S. Pyne is with (1) Public Health Dynamics Lab, and Department of Biostatistics, Graduate School of Public Health, University of Pittsburgh, Pittsburgh, Pennsylvania, USA.
(2) Health Analytics Network, Pennsylvania, USA., e-mail: {\tt spyne@pitt.edu}}
 }

\date{}
 
\maketitle

\begin{abstract}
Social distancing and stay-at-home are among the few measures that are known to be effective in checking the spread of a pandemic such as COVID-19 in a given population. The patterns of dependency between such measures and their effects on disease incidence may vary dynamically and across different populations. We described a new computational framework to measure and compare the temporal relationships between human mobility and new cases of COVID-19 across more than 150 cities of the United States with relatively high incidence of the disease. We used a novel application of Optimal Transport for computing the distance between the normalized patterns induced by bivariate time series for each pair of cities. Thus, we identified 10 clusters of cities with similar temporal dependencies, and computed the Wasserstein barycenter to describe the overall dynamic pattern for each cluster. Finally, we used city-specific socioeconomic covariates to analyze the composition of each cluster.  
\end{abstract}


\section{Introduction}
The phenomenal spread of the COVID-19 pandemic will have unprecedented consequences for human life and livelihood. In the absence of a treatment or vaccine to develop immunity against the disease, governments around the world have used non-pharmaceutical, risk mitigation strategies such as lockdowns, shelter-in-place, school and business closures, travel bans or restrictions to limit movement and prevent contagion. The magnitude and effectiveness of such mitigation strategies in preventing contagion and reducing the number of deaths is shown in Europe where such mitigation strategies have reduced the reproduction number over time $(R_t)$ below 1, which means that the virus will gradually stop spreading. Since the beginning of the epidemic, an estimated 3.1 million deaths were averted across 11 European countries attributable to these risk mitigation strategies \cite{flaxman2020estimating}. 

In the United States, the adoption, and enforcement of non-pharmaceutical, risk mitigation strategies have varied by state and across time. The first confirmed COVID-19 case was reported on January 21, 2020, in Washington State \cite{ghinai2020first}. While transmissions were documented since, a national emergency was declared later on March 13 \cite{house2020proclamation}. At that time, international travel restrictions were enforced. By March 16, six bay area counties declared shelter-in-place orders and on March 19, California was the first state to issue a state-wide order. Since then, several communities and states implemented stay-at-home orders and social distancing measures. As of March 30, there were 162,600 confirmed COVID-19 cases in the U.S. \cite{house2020proclamation} and 30 states had announced shelter-in-place orders. On April 1, two additional states and the District of Columbia issued statewide shelter-in-place orders followed by 7 more states by April 6. 

Historically, among the U.S. cities that were hit by the 1918 Spanish ﬂu, social distancing played a pivotal role in ﬂattening the pandemic curve. In fact, the cities which delayed enforcing social distancing saw the highest peaks in new cases of the disease. Policies aimed at reducing human transmission of COVID-19 included lockdown, travel restrictions, quarantine, curfew, cancellation and postponing events, and facility closures. Measuring the dynamic impact of these interventions is challenging \cite{adiga2020interplay,dascritical} and confounded by several factors such as differences in the specific modes and dates of the policy-driven measures adopted by or enforced across states, regions, and countries, and, of course, the actual diversity of human behaviors at these locations. 

Given the current ubiquitous usage of mobile devices among the U.S. populations, social mobility as measured by aggregating the geospatial statistics of their daily movements could serve as a proxy measure to assess the impact of such policies as social distancing on human transmission. In the particular context of the current pandemic, human mobility data could be estimated using geolocation reports from user smartphones and other mobile devices that were made available by multiple providers including Google and Apple, among others. In this study, we obtained such data from Descartes Labs, which made anonymized location-specific time series data on mobility index freely available to researchers through their GitHub site: \url{https://github.com/descarteslabs/DL-COVID-19.} Thus, we obtained location-specific bivariate time series on daily mobility index and disease incidence, i.e., new cases of COVID-19 in the U.S.
 
In this study, we are interested to (a) measure and compare the temporal dependencies between mobility ($M$) and new cases ($N$) across 151 cities in the U.S. with relatively high incidence of COVID-19 by May 31, 2010. We believe that these dependency patterns vary not only over time but across locations and populations. For this purpose, we proposed a novel application of Optimal Transport to compute the distance between patterns of ($N$, mobility, time) and its variants for each pair of cities. This allowed us to (b) group the cities into different hierarchical clusterings, and (c) compute the barycenter to describe the overall dynamic pattern of each identified cluster. Finally, we also used city-specific socioeconomic covariates to analyze the composition of each cluster. A pipeline for our analytical framework is described in the following section.

\begin{figure}
    \centering
    \includegraphics[scale=0.2]{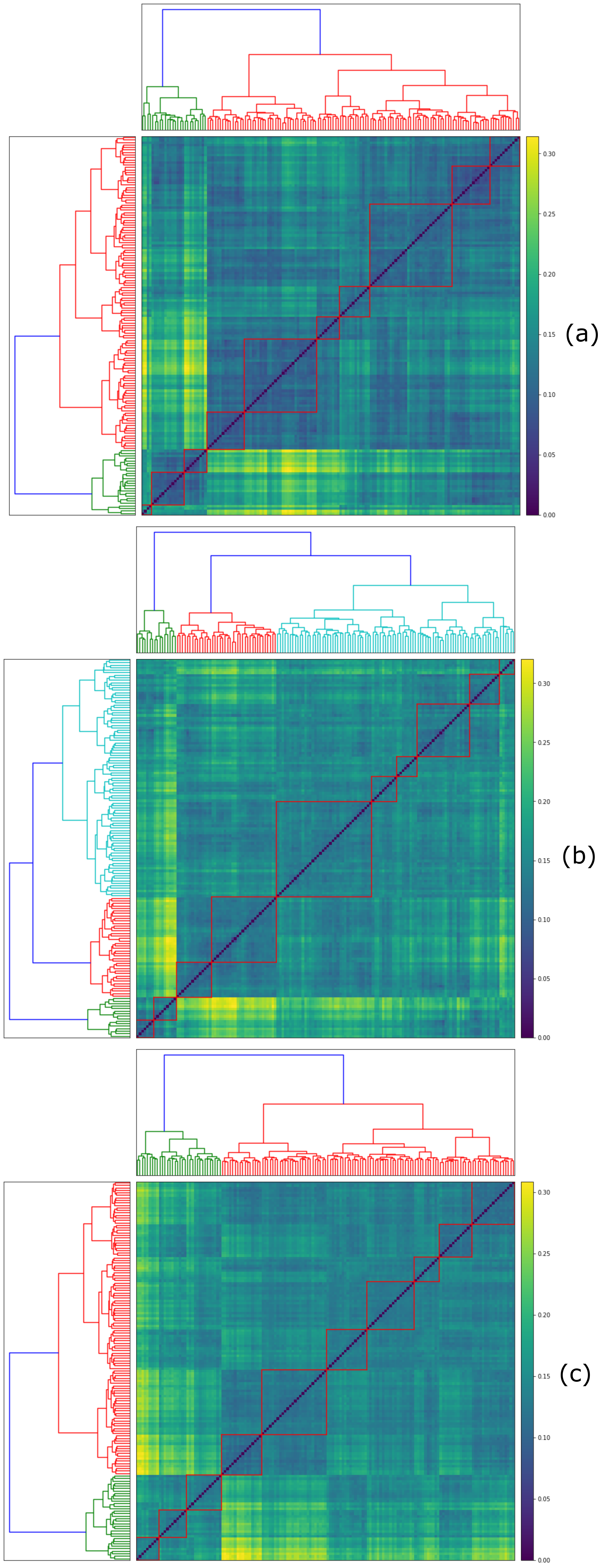}
    \caption{The dendrograms show 3 hierarchical clusterings of cities (a), (b) and (c) respectively based on ($N$, $M$, $t$), ($N$, $\Delta M$, $t$) and ($N$, $M'$, $t$) using Ward's linkage. Based on visual inspection of the seriated distance matrix, 10 clusters were identified in each case, as shown on the heatmaps.}
    \label{fig:f1}
\end{figure}

\section{Data and Methods}

\subsection{Datasets}
\subsubsection{COVID-19 incidence and population data}
    Based on cumulative  COVID-19 cases data from the Johns Hopkins Coronavirus Resource Center (\url{https://coronavirus.jhu.edu/}), for this study, we compiled time series data on daily new cases of the disease for more than 300 U.S. counties from 32 states and the District of Columbia and matched by five-digit FIPS code or county name to dynamic and static variables from additional data sources. Since a single county may consist of multiple individual cities, we include the list of all city labels within each aggregate group to represent a greater metropolitan area. A total of 151 of such metropolitan areas that had at least 1,000 reported cases of COVID-19 by May 31, 2020, were selected for this study. Population covariates for these areas were collected from the online resources of the U.S. Census Bureau and the U.S. Centers for Disease Control and Prevention (CDC)  (\url{https://www.census.gov/quickfacts/}, \url{https://svi.cdc.gov/}).

\subsubsection{Human mobility index data}
Anonymized geolocated mobile phone data from several providers including Google and Apple, timestamped with local time, were recently made available for analysis of human mobility patterns during the pandemic. Based on geolocation pings from a collection of mobile devices reporting consistently throughout the day, anonymous aggregated mobility indices were calculated for each county at Descartes Lab. The maximum distance moved by each node, after excluding outliers, from the first reported location was calculated. Using this value, the median across all devices in the sample is computed to generate a mobility metric for select locations at county level. Descartes Labs further defines a normalized mobility index as a proportion of the median of the maximum distance mobility to the ``normal'' median during an earlier time-period multiplied by a factor of 100. Thus, the mobility index provides a baseline comparison to evaluate relative changes in population behavior during COVID-19 pandemic.\cite{warren2020mobility}.

\begin{figure}
    \centering
    \includegraphics[scale=0.6]{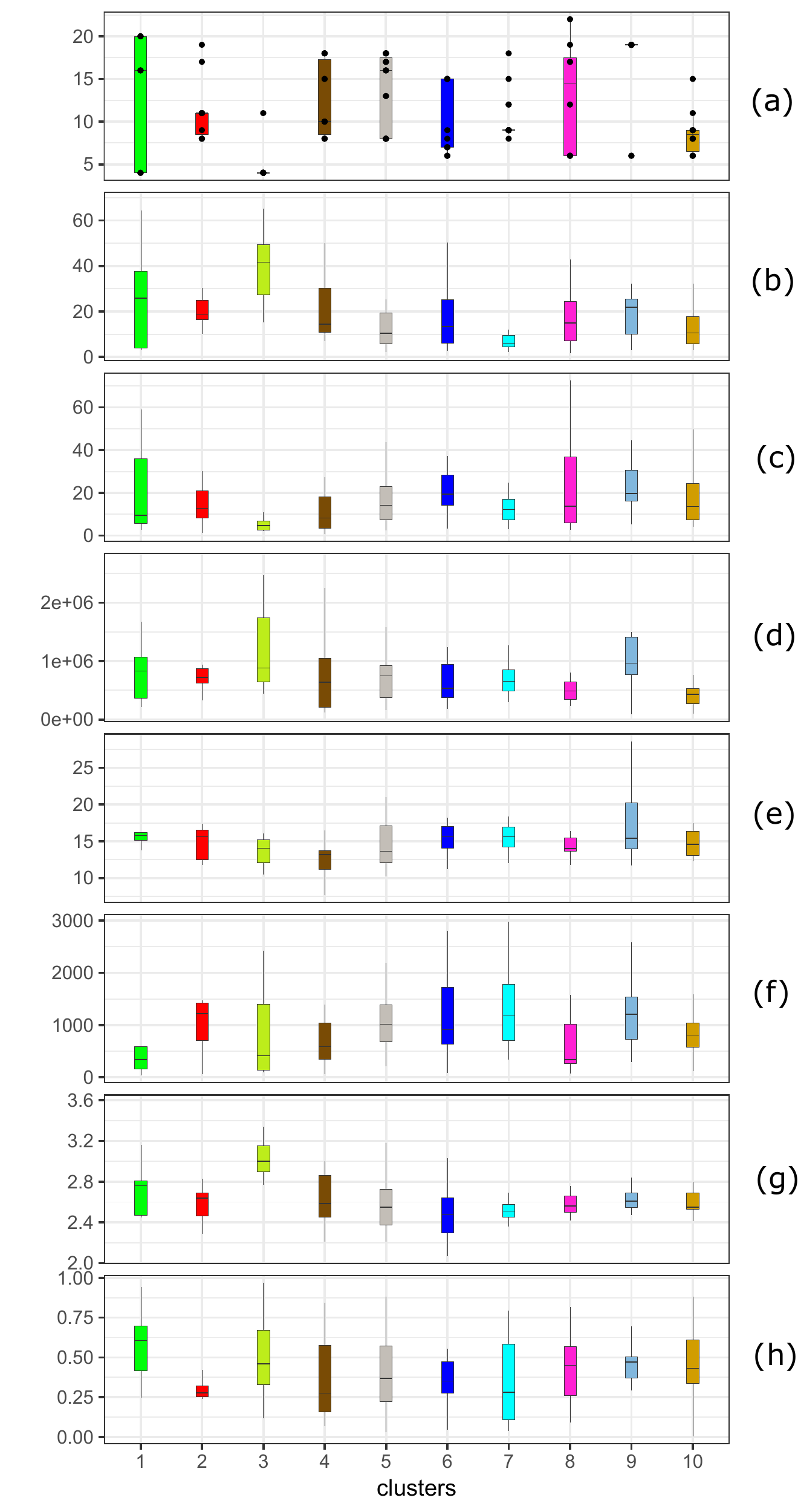}
    \caption{The boxplots show the differences across the identified 10 clusters of cities in terms of the values of the 8 most significant covariates: (a) Reaction Time (RT), (b) hispanic percent, (c) black percent, (d) population size, (e) senior percent, (f) population density 2010, (g) persons per household, and (h) SVI ses. We jittered the overlapping RT points for easy visualization.}
    \label{fig:f6}
\end{figure}

\subsection{Methods}
Below we list the steps of the overall workflow of our framework, and briefly describe the same in the following paragraphs of this section. 

\begin{algorithm}
    \caption{The workflow of the analytical framework}
    \begin{algorithmic}[1]
    \renewcommand{\algorithmicrequire}{\textbf{Input:}} 
   \renewcommand{\algorithmicensure}{\textbf{Steps of the Analysis:}}
    \REQUIRE {For each of $k (=151)$ given cities, a bivariate time series: mobility ($M$) and new cases ($N$) for each date ($t$) over a fixed time-interval (March 1 -- May 31, 2020).}
   \ENSURE .
   % \textbf{Steps}:
    
    \STATE As measures of mobility, along with $M$, also consider its variants $\Delta M$ and $M'$ computed with equations \ref{eq1} and \ref{eq2}. 
    \STATE Performed normalized ranking of variables ($M$/$\Delta M$/$M'$, $N$ and $t$) to represent each city as a discrete set of ranked points in unit cube ($[0, 1]^3$)  
    \STATE Compute optimal transport (OT) distance between the pointsets representing each pair of cities. 
    \STATE Cluster the cities based on the OT distance matrix. Three different hierarchical clusterings $HC1$, $HC2$ and $HC3$ were obtained based on Ward's linkage method and 3 variants of mobility: $M$, $\Delta M$, and $M'$ respectively.
    \STATE Apply HCMapper to compare the dendrograms of different clusterings ($HC1$, $HC2$ and $HC3$). Select the clustering ($HC3$) that yields the most spatially homogeneous clusters. 
    \STATE Compute Wasserstein barycenter for each cluster of the selected clustering ($HC3$). 
    \STATE Analyze the composition of the clusters by applying random forest classifier on 15 city-specific covariates as feature set. Identify the contributions of the covariates to discriminate among the clusters.  
    %    \label{algo:cbfs}
    \end{algorithmic}
\end{algorithm}

\subsubsection{Temporal patterns of mobility}
To better understand the temporal patterns of mobility, in addition to the given non-negative mobility index $M$, we also use two variants: delta mobility $\Delta M$ and $M'$ defined as follows:
\begin{equation}
\Delta M(t)= M(t)-M(t-1)\\
\label{eq1}
\end{equation}
and
\begin{equation}
  M'(t)=((M(t)-M(t-1))+0.5(M(t+1)-M(t-1)))/2.
\label{eq2}
\end{equation}
 Here $\Delta M$ is the first difference, and $M'$ approximately the local derivative \cite{keogh2001derivative}, of the time series $M$, and yet, unlike $M$, these are not restricted to be non-negative.
 
 \subsubsection{Representing a city as discrete set of points} With the above definitions, the temporal relationship between mobility (and its variants) and new cases of each city in our data can be depicted as tuples ($M/\Delta M/M'$, $N$, $t$). We represent the time series by performing a normalized ranking of the variables so as to represent each city by a discrete set of points in unit cube $[0, 1]^3$. This normalized ranking is frequently used as a estimator for empirical copulas with good convergence properties \cite{deheuvels1980non}. The cities can have different representations by considering the three definitions of mobility metrics, and in each case, we can have different groupings of cities. A comparative analysis of all groupings can provide a correlation structure between groups of cities from different perspectives.

\subsubsection{Comparing cities using optimal transport}
To distinguish between the temporal dependence between mobility and new cases of two cities, we use Wasserstein distance from optimal transport theory. We compute Wasserstein distance between two discrete sets of points in unit cube, corresponding to two cities, as the minimum cost of transforming the discrete distribution of one set of points to the other set. It can be computed without the need of such steps as fitting kernel densities or arbitrary binning that can introduce noise to data. Wasserstein distance between two distributions on a given metric space $M$ is conceptualized by the minimum ``cost" to transport or morph one pile of dirt into another -- the so-called `earth mover's distance'. This ``global'' minimization over all possible ways to morph takes into consideration the ``local'' cost of morphing each grain of dirt across the piles \cite{peyre2019computational}.

Given a metric space $\calM$, the distance optimally transports the probability $\mu$ defined over $\calM$ to turn it into $\nu$:
\begin{equation}
    W_p(\mu,\nu)=\left(\inf_{\lambda \in \tau(\mu,\nu)} \int _{\calM \times \calM} d(x,y)^p d\lambda (x,y)\right)^{1/p},
\end{equation} where $p \ge 1$, $\tau(\mu,\nu)$ denotes the collection of all measures on $\calM\times \calM$ with marginals $\mu$ and $\nu$. The intuition and motivation of this metric came from optimal transport problem, a classical problem in mathematics, which was first introduced by the French mathematician Gaspard Monge in 1781 and later formalized in a relaxed form by L.~Kantorovitch in 1942.

\subsubsection{Clustering the cities}
Upon computing optimal transport based distances for each pair of cities, hierarchical clustering of the cities is performed using Ward's minimum variance method \cite{inbook}. For the 3 variants of mobility ($M/\Delta M/M'$), we obtained 3 different hierarchical clusterings: $\HC{}1$, $\HC{}2$ and $\HC{}3$ respectively. Based on visual inspection of the distance matrix seriated by the hierarchical clustering, and looping over the number of clusters, we take a relevant flat cut in the dendrogram. For each case, we got 10 clusters, each consisting of cities that are similar with respect to their dependence between mobility and new cases.

\subsubsection{Comparing the clusterings}
The resulting clusters are compared using a visualization tool called HCMapper \cite{marti2015hcmapper}. HCMapper can compare a pair of dendrograms of two different hierarchical clusterings computed on the same dataset. It aims to find clustering singularities between two models by displaying multiscale partition-based layered structures. The three different clustering results are compared with HCMapper to sought out the structural instabilities of clustering hierarchies. In particular, the display graph of HCMapper has $n$ columns, where $n$ represents the number of hierarchies we want to compare (here $n=3$). Each column consists of the same number of flat clusters, which is depicted as rectangles within the column. Rectangle size is proportional to the number of cities within the clusters, while an edge between two clusters tells the number of shared cities between them. Thus, a one-to-one mapping between the clusters of two columns preferably depicts a similar perfect clustering while too many edges crossing between two columns describe a dissimilar structure. 

We also checked the spatial homogeneity of a clustering in terms of the average number of clusters in which the cities of each state were assigned to, over all states that are represented in our data. Moran's $I$ to assess the spatial correlation among the cluster labels was also computed. 

\subsubsection{Summarizing the distinctive cluster patterns}
We summarize the overall pattern of each identified cluster by computing its barycenter in Wasserstein space. It efficiently describes the underlying temporal dependence between the measures of mobility (here we use $M'$) and incidence within each cluster. 
Wasserstein distances have several important theoretical and practical properties \cite{villani2008optimal, pele2009fast}. Among these, a barycenter in Wasserstein space is an appealing concept which already shows a high potential in different applications such as, in artificial intelligence, machine learning and Statistics \cite{carlier2015numerical,le2017existence,benamou2015iterative,cuturi2014fast}.  

A Wasserstein barycenter \cite{agueh2011barycenters, cuturi2014fast} of $n$ measures $\nu_1 \ldots \nu_n$ in $\mathbb{P} \in P(\calM)$ is defined as a minimizer of the function $f$ over $\mathbb{P}$, where
\begin{equation}
    f(\mu)=\frac{1}{N}\sum_{i=1}^N W_p^p(\nu_i,\mu).
\end{equation}

A fast algorithm \cite{cuturi2014fast} was proposed to minimize the sum of optimal transport distances from one measure (the variable) to a set of fixed measures using gradient descent. These gradients are computed using matrix scaling algorithms in a considerable lower computational cost. We have used the method proposed in \cite{cuturi2014fast} and implemented in the POT library (\url{https://pythonot.github.io/}) to compute the barycenter of each cluster.

\subsubsection{Analysis of the clusters using static covariates}
To understand the composition of the identified clusters, i.e., what could explain the similarity in the temporal dependence between mobility and new cases of the cities that belong to a cluster, we used different city-specific population covariates, while checking their relative contributions to discriminating the clusters. These covariates include  (a) date of Stay-at-home order, (b) population	size, (c) persons per household,	(d) senior percentage, (e)	black percent, (f) hispanic percent,	(g) poor percent,	(h) population density 2010, (i) SVI ses (j) SVI minority, (k) SVI overall, and (l) Gini index. Here SVI stands for Social Vulnerability Index of CDC, and ``ses" socioeconomic status. In addition, we also compute the `reaction time' (RT) of each city as the number of days between the stay-at-home-order at a given city and a common reference stating point date (taken as 15 March, 2020). 

This step also provides a form of external validation of the clustering results as none of the above covariates were used for clustering. To demonstrate, we conducted this step with the clustering $\HC{}3$ obtained from the time series $M'$.

Using the covariates as features of the cities, a random forest classifier is trained to learn the cluster labels. The aim is to see how the clustering could be explained by the covariates. To find which of the features contribute most to discriminate the clusters of cities we computed the mean Shapley values \cite{NIPS2017_7062}. A Shapley value quantifies the magnitude of the impact of the features on the classification task. The ranking of the covariates/features based on the mean Shapley values determines the most relevant features in this regard.

\section{Results}
\begin{figure}
    \centering
    \includegraphics[scale=0.6]{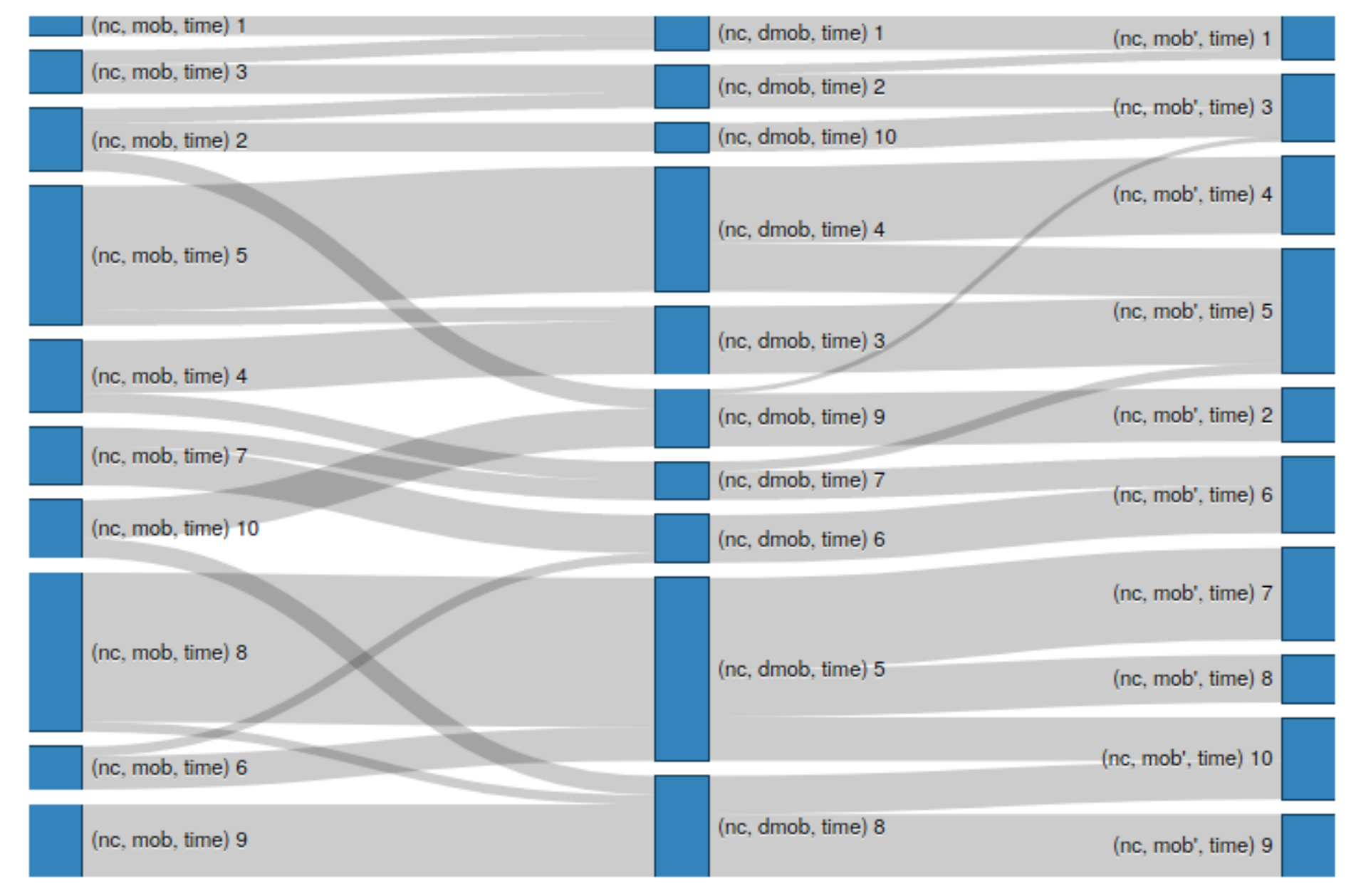}
    \caption{HCMapper is used for comparison of 3 hierarchical clusterings of cities based on $\HC{}1$($N$, $M$, $t$), $\HC{}2$($N$, $\Delta M$, $t$) and $\HC{}3$($N$, $M'$, $t$). The cluster sizes and divergences across the clusterings are shown with blue rectangles and grey edges respectively.}
    \label{fig:f2}
\end{figure}

In this study, we used bivariate time series on daily values of mobility index and COVID-19 incidence over a 3-month time-period (March 1 -- May 31, 2020) for 151 U.S. cities that have reported at least 1,000 cases by the end of this period. By transforming the data for each city to a corresponding discrete set of ranked points on the unit cube, we computed the Optimal Transport distance as measure of temporal dependency between mobility and new cases for each pair of cities. Three definitions of mobility ($M$/$\Delta M$/$M'$) allowed us to generate 3 hierarchical clusterings: $HC1$, $HC2$ and $HC3$, as shown in Figure \ref{fig:f1} and Table \ref{longtab} . Each of the clusterings yielded 10 clusters of cities, which were compared for their sizes, singularities and divergences by the tool HCMapper, as shown in Figure \ref{fig:f2}.

\begin{figure}
    \centering
    \includegraphics[scale=2.3]{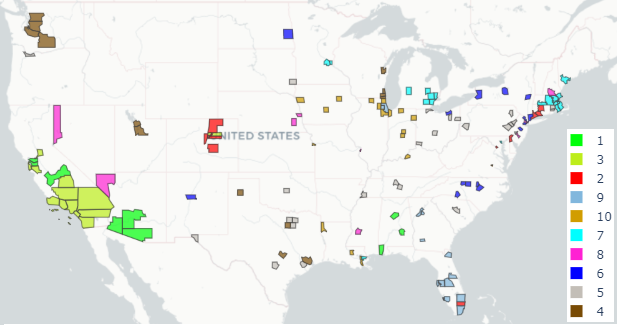}
    \caption{The geographic distribution of the 10 identified clusters by $HC3$ are shown. The county corresponding to each city is mapped in its cluster-specific color.}
    \label{fig:f3}
\end{figure}

Among the clusterings, $HC3$ appeared to have clusters of consistent sizes, and also the fewest singularities and divergences. Further, when we mapped the counties representing the cities with cluster-specific colors, as shown in Figure \ref{fig:f3}, we observed that the $HC3$ clusters showed high spatial correlation (Moran's $I$ p-value of 0). They also showed the least disagreements among the cluster assignments of cities with each state, although some states like California and Florida contained cities from more than one cluster (see table \ref{longtab}). We looked into possible explanations of such cluster-specific differences using local covariates, as described below.

\begin{figure}
    \centering
    \includegraphics[scale=0.25]{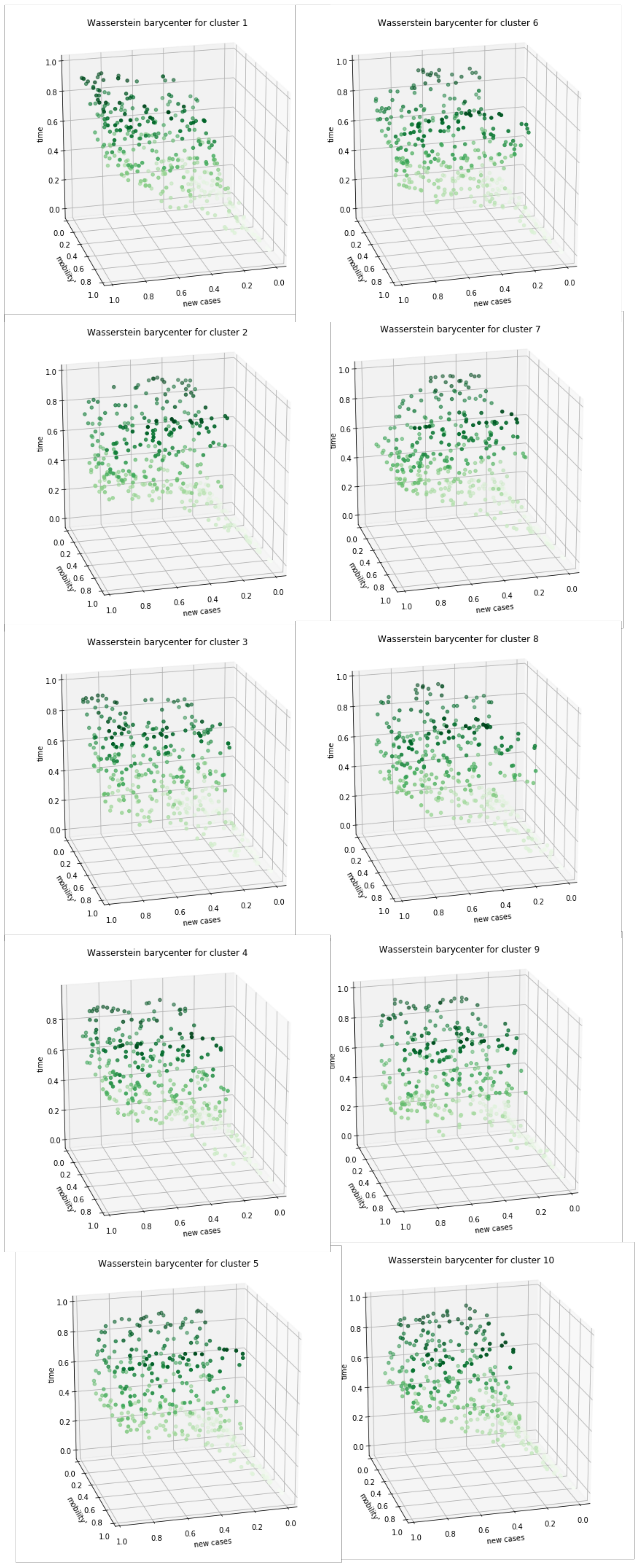}
    \caption{The overall temporal pattern of dependency between normalized measures of mobility and COVID-19 incidence for each identified cluster of cities is shown along 3-dimensions ($N$, $M'$, $t$). The Wasserstein barycenters of the 10 clusters are depicted within the unit cube with the darker dots representing later points in time (z-axis).}
    \label{fig:f4}
\end{figure}

Given the assumption of this study is that there are dynamic relationships between mobility and COVID-19 incidence that change not only over time but also across locations and populations, we computed Wasserstein barycenters of the 10 identified clusters, as shown in Figure \ref{fig:f4}, to describe the overall dependency structure that is specific to each cluster. The temporal changes in the dependencies are shown in 3-dimensional plots, as the shading changes from light (early points) to dark green (later points) along the z-axis (time).

\begin{figure}
    \centering
    \includegraphics[scale=0.6]{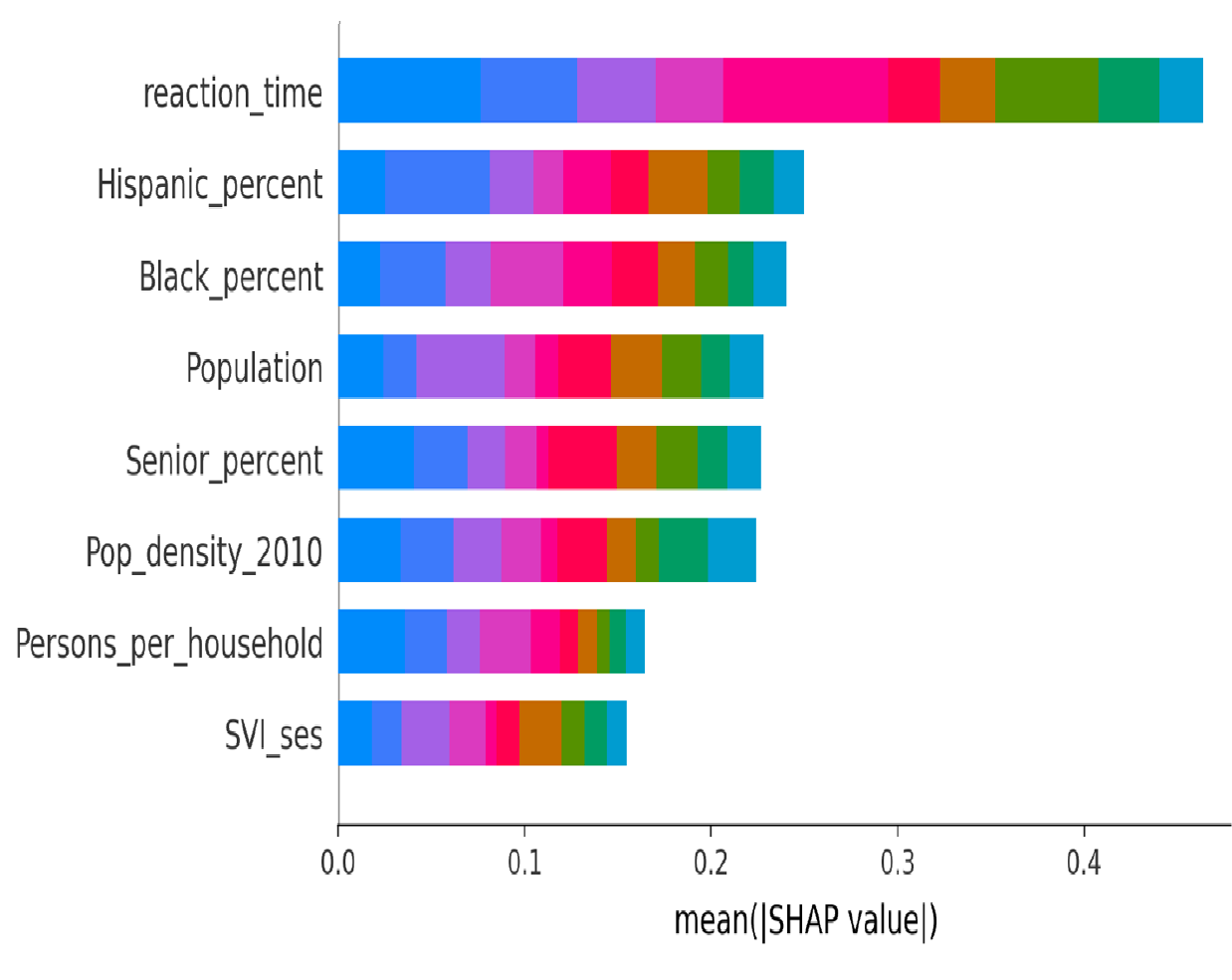}
    \caption{The relative contributions of 8 most significant static city-specific covariates in discrimination of the 10 clusters identified by $\HC{}3$ and shown with different colors. The contributions towards each cluster are measured by mean Shapley values for each covariate.}
    \label{fig:f5}
\end{figure}

Finally, we sought to understand the factors that possibly underlie the dynamic patterns of each cluster as described above. Towards this, our results from Random Forest classification identified socioeconomic characteristics (or covariates) of the cities that could discriminate among the assigned cluster labels. The 8 most significantly discriminating covariates are shown in Figure \ref{fig:f5} along with their cluster-specific contributions measured by the mean Shapley values. Notably, none of these covariates were used for clustering, and are yet able to discriminate among the clusters. Figure \ref{fig:f6} shows the distinctive distributions of these covariates across the 10 identified clusters as boxplots. Reaction time is robustly the first and major contributor, which is indicative of the effects of stay-at-home on the different patterns of COVID-19 dynamics.

\vspace{0.1in}

{
\topcaption{Table of 151 cities with their respective Date (mm.dd.2020) of stay-at-home order, Reaction Time (RT), and clusters labels assigned by HC1, HC2 and HC3. The absence of stay-at-home order is denoted by NA.}%frank\\
\label{longtab}
\centering
\begin{supertabular}{|p{1.3in}|p{0.5in}|p{0.2in}|p{0.2in}|p{0.2in}|p{0.2in}|}\hline
County & Date & RT & HC1 & HC2 & HC3 \\ \hline
Jefferson, AL & 4.4 & 20 & 1 & 1 & 1 \\ \hline
Mobile, AL & 4.4 & 20 & 1 & 1 & 1 \\ \hline
Montgomery, AL & 4.4 & 20 & 1 & 1 & 1 \\ \hline
Maricopa, AZ & 3.31 & 16 & 1 & 1 & 1 \\ \hline
Pima, AZ & 3.31 & 16 & 3 & 1 & 1 \\ \hline
Yuma, AZ & 3.31 & 16 & 3 & 1 & 1 \\ \hline
Alameda, CA & 3.19 & 4 & 3 & 1 & 1 \\ \hline
Contra Costa, CA & 3.19 & 4 & 3 & 2 & 1 \\ \hline
Fresno, CA & 3.19 & 4 & 3 & 2 & 1 \\ \hline
Kern, CA & 3.19 & 4 & 3 & 2 & 3 \\ \hline
Los Angeles, CA & 3.19 & 4 & 3 & 2 & 3 \\ \hline
Orange, CA & 3.19 & 4 & 3 & 2 & 3 \\ \hline
Riverside, CA & 3.19 & 4 & 3 & 2 & 3 \\ \hline
Sacramento, CA & 3.19 & 4 & 2 & 2 & 3 \\ \hline
San Bernardino, CA & 3.19 & 4 & 2 & 2 & 3 \\ \hline
San Diego, CA & 3.19 & 4 & 2 & 2 & 3 \\ \hline
San Francisco, CA & 3.19 & 4 & 2 & 10 & 3 \\ \hline
San Mateo, CA & 3.19 & 4 & 2 & 10 & 3 \\ \hline
Santa Barbara, CA & 3.19 & 4 & 2 & 10 & 3 \\ \hline
Santa Clara, CA & 3.19 & 4 & 2 & 10 & 3 \\ \hline
Tulare, CA & 3.19 & 4 & 2 & 10 & 3 \\ \hline
Ventura, CA & 3.19 & 4 & 2 & 10 & 3 \\ \hline
Adams, CO & 3.26 & 11 & 2 & 9 & 3 \\ \hline
Arapahoe, CO & 3.26 & 11 & 2 & 9 & 2 \\ \hline
Denver, CO & 3.26 & 11 & 2 & 9 & 2 \\ \hline
El Paso, CO & 3.26 & 11 & 2 & 9 & 2 \\ \hline
Jefferson, CO & 3.26 & 11 & 10 & 9 & 2 \\ \hline
Weld, CO & 3.26 & 11 & 10 & 9 & 2 \\ \hline
Fairfield, CT & 3.23 & 8 & 10 & 9 & 2 \\ \hline
Hartford, CT & 3.23 & 8 & 10 & 9 & 2 \\ \hline
New Haven, CT & 3.23 & 8 & 10 & 9 & 2 \\ \hline
New Castle, DE & 3.24 & 9 & 10 & 9 & 2 \\ \hline
Washington, DC & 4.1 & 17 & 10 & 9 & 2 \\ \hline
Broward, FL & 4.3 & 19 & 10 & 9 & 2 \\ \hline
Duval, FL & 4.3 & 19 & 10 & 8 & 9 \\ \hline
Hillsborough, FL & 4.3 & 19 & 10 & 8 & 9 \\ \hline
Lee, FL & 4.3 & 19 & 10 & 8 & 9 \\ \hline
Miami-Dade, FL & 4.3 & 19 & 10 & 8 & 9 \\ \hline
Orange, FL & 4.3 & 19 & 9 & 8 & 9 \\ \hline
Palm Beach, FL & 4.3 & 19 & 9 & 8 & 9 \\ \hline
Pinellas, FL & 4.3 & 19 & 9 & 8 & 9 \\ \hline
Polk, FL & 4.3 & 19 & 9 & 8 & 9 \\ \hline
DeKalb, GA & 4.3 & 19 & 9 & 8 & 9 \\ \hline
Dougherty, GA & 4.3 & 19 & 9 & 8 & 9 \\ \hline
Fulton, GA & 4.3 & 19 & 9 & 8 & 9 \\ \hline
Cook, IL & 3.21 & 6 & 9 & 8 & 9 \\ \hline
DuPage, IL & 3.21 & 6 & 9 & 8 & 9 \\ \hline
Kane, IL & 3.21 & 6 & 9 & 8 & 10 \\ \hline
Lake, IL & 3.21 & 6 & 9 & 8 & 10 \\ \hline
Will, IL & 3.21 & 6 & 9 & 8 & 10 \\ \hline
Winnebago, IL & 3.21 & 6 & 9 & 8 & 10 \\ \hline
Allen, IN & 3.24 & 9 & 9 & 8 & 10 \\ \hline
Hamilton, IN & 3.24 & 9 & 9 & 8 & 10 \\ \hline
Lake, IN & 3.24 & 9 & 8 & 8 & 10 \\ \hline
Marion, IN & 3.24 & 9 & 8 & 8 & 10 \\ \hline
St. Joseph, IN & 3.24 & 9 & 8 & 5 & 10 \\ \hline
Black Hawk, IA & NA & 85 & 8 & 5 & 10 \\ \hline
Polk, IA & NA & 85 & 8 & 5 & 10 \\ \hline
Woodbury, IA & NA & 85 & 8 & 5 & 10 \\ \hline
Wyandotte, KS & 3.3 & 15 & 8 & 5 & 10 \\ \hline
Jefferson, KY & 3.26 & 11 & 8 & 5 & 10 \\ \hline
Caddo, LA & 3.23 & 8 & 8 & 5 & 10 \\ \hline
East Baton Rouge, LA & 3.23 & 8 & 8 & 5 & 10 \\ \hline
Jefferson, LA & 3.23 & 8 & 8 & 5 & 10 \\ \hline
Orleans, LA & 3.23 & 8 & 8 & 5 & 7 \\ \hline
Cumberland, ME & 4.2 & 18 & 8 & 5 & 7 \\ \hline
Baltimore City, MD & 3.3 & 15 & 8 & 5 & 7 \\ \hline
Bristol, MA & 3.24 & 9 & 8 & 5 & 7 \\ \hline
Essex, MA & 3.24 & 9 & 8 & 5 & 7 \\ \hline
Hampden, MA & 3.24 & 9 & 8 & 5 & 7 \\ \hline
Middlesex, MA & 3.24 & 9 & 8 & 5 & 7 \\ \hline
Norfolk, MA & 3.24 & 9 & 8 & 5 & 7 \\ \hline
Plymouth, MA & 3.24 & 9 & 8 & 5 & 7 \\ \hline
Suffolk, MA & 3.24 & 9 & 8 & 5 & 7 \\ \hline
Worcester, MA & 3.24 & 9 & 8 & 5 & 7 \\ \hline
Genesee, MI & 3.24 & 9 & 8 & 5 & 7 \\ \hline
Kent, MI & 3.24 & 9 & 8 & 5 & 7 \\ \hline
Macomb, MI & 3.24 & 9 & 8 & 5 & 7 \\ \hline
Oakland, MI & 3.24 & 9 & 8 & 5 & 7 \\ \hline
Washtenaw, MI & 3.24 & 9 & 8 & 5 & 7 \\ \hline
Wayne, MI & 3.24 & 9 & 8 & 5 & 7 \\ \hline
Hennepin, MN & 3.27 & 12 & 8 & 5 & 7 \\ \hline
Ramsey, MN & 3.27 & 12 & 8 & 5 & 7 \\ \hline
Hinds, MS & 4.3 & 19 & 8 & 5 & 8 \\ \hline
St. Louis City, MO & 4.6 & 22 & 8 & 5 & 8 \\ \hline
Douglas, NE & NA & 85 & 8 & 5 & 8 \\ \hline
Lancaster, NE & NA & 85 & 6 & 5 & 8 \\ \hline
Clark, NV & 4.1 & 17 & 6 & 5 & 8 \\ \hline
Washoe, NV & 4.1 & 17 & 6 & 5 & 8 \\ \hline
Hillsborough, NH & 3.27 & 12 & 6 & 5 & 8 \\ \hline
Camden, NJ & 3.21 & 6 & 6 & 5 & 8 \\ \hline
Essex, NJ & 3.21 & 6 & 6 & 5 & 8 \\ \hline
Hudson, NJ & 3.21 & 6 & 6 & 5 & 8 \\ \hline
Mercer, NJ & 3.21 & 6 & 6 & 6 & 6 \\ \hline
Passaic, NJ & 3.21 & 6 & 6 & 6 & 6 \\ \hline
Union, NJ & 3.21 & 6 & 7 & 6 & 6 \\ \hline
Bernalillo, NM & 3.24 & 9 & 7 & 6 & 6 \\ \hline
Albany, NY & 3.22 & 7 & 7 & 6 & 6 \\ \hline
Erie, NY & 3.22 & 7 & 7 & 6 & 6 \\ \hline
New York City, NY & 3.22 & 7 & 7 & 6 & 6 \\ \hline
Onondaga, NY & 3.22 & 7 & 7 & 6 & 6 \\ \hline
Westchester, NY & 3.22 & 7 & 7 & 6 & 6 \\ \hline
Durham, NC & 3.3 & 15 & 7 & 6 & 6 \\ \hline
Forsyth, NC & 3.3 & 15 & 7 & 7 & 6 \\ \hline
Guilford, NC & 3.3 & 15 & 7 & 7 & 6 \\ \hline
Mecklenburg, NC & 3.3 & 15 & 7 & 7 & 6 \\ \hline
Wake, NC & 3.3 & 15 & 7 & 7 & 6 \\ \hline
Cass, ND & NA & 85 & 4 & 7 & 6 \\ \hline
Cuyahoga, OH & 3.23 & 8 & 4 & 7 & 6 \\ \hline
Franklin, OH & 3.23 & 8 & 4 & 7 & 5 \\ \hline
Hamilton, OH & 3.23 & 8 & 4 & 7 & 5 \\ \hline
Lucas, OH & 3.23 & 8 & 4 & 3 & 5 \\ \hline
Mahoning, OH & 3.23 & 8 & 4 & 3 & 5 \\ \hline
Summit, OH & 3.23 & 8 & 4 & 3 & 5 \\ \hline
Oklahoma, OK & NA & 85 & 4 & 3 & 5 \\ \hline
Multnomah, OR & 3.23 & 8 & 4 & 3 & 5 \\ \hline
Allegheny, PA & 3.23 & 8 & 4 & 3 & 5 \\ \hline
Berks, PA & 4.1 & 17 & 4 & 3 & 5 \\ \hline
Lackawanna, PA & 4.1 & 17 & 4 & 3 & 5 \\ \hline
Lehigh, PA & 4.1 & 17 & 4 & 3 & 5 \\ \hline
Northampton, PA & 4.1 & 17 & 4 & 3 & 5 \\ \hline
Philadelphia, PA & 3.23 & 8 & 4 & 3 & 5 \\ \hline
Kent, RI & 3.28 & 13 & 5 & 3 & 5 \\ \hline
Providence, RI & 3.28 & 13 & 5 & 3 & 5 \\ \hline
Richland, SC & NA & 85 & 5 & 3 & 5 \\ \hline
Minnehaha, SD & NA & 85 & 5 & 4 & 5 \\ \hline
Davidson, TN & 3.31 & 16 & 5 & 4 & 5 \\ \hline
Rutherford, TN & 3.31 & 16 & 5 & 4 & 5 \\ \hline
Shelby, TN & 3.31 & 16 & 5 & 4 & 5 \\ \hline
Bexar, TX & 4.2 & 18 & 5 & 4 & 5 \\ \hline
Collin, TX & 4.2 & 18 & 5 & 4 & 5 \\ \hline
Dallas, TX & 4.2 & 18 & 5 & 4 & 5 \\ \hline
Denton, TX & 4.2 & 18 & 5 & 4 & 5 \\ \hline
El Paso, TX & 4.2 & 18 & 5 & 4 & 5 \\ \hline
Fort Bend, TX & 4.2 & 18 & 5 & 4 & 5 \\ \hline
Harris, TX & 4.2 & 18 & 5 & 4 & 4 \\ \hline
Potter, TX & 4.2 & 18 & 5 & 4 & 4 \\ \hline
Tarrant, TX & 4.2 & 18 & 5 & 4 & 4 \\ \hline
Travis, TX & 4.2 & 18 & 5 & 4 & 4 \\ \hline
Salt Lake, UT & NA & 85 & 5 & 4 & 4 \\ \hline
Utah, UT & NA & 85 & 5 & 4 & 4 \\ \hline
Alexandria, VA & 3.3 & 15 & 5 & 4 & 4 \\ \hline
Richmond City, VA & 3.3 & 15 & 5 & 4 & 4 \\ \hline
King, WA & 3.23 & 8 & 5 & 4 & 4 \\ \hline
Pierce, WA & 3.23 & 8 & 5 & 4 & 4 \\ \hline
Snohomish, WA & 3.23 & 8 & 5 & 4 & 4 \\ \hline
Yakima, WA & 3.23 & 8 & 5 & 4 & 4 \\ \hline
Brown, WI & 3.25 & 10 & 5 & 4 & 4 \\ \hline
Kenosha, WI & 3.25 & 10 & 5 & 4 & 4 \\ \hline
Milwaukee, WI & 3.25 & 10 & 5 & 4 & 4 \\ \hline
Racine, WI & 3.25 & 10 & 5 & 4 & 4 \\ \hline
%\label{longtab}
\end{supertabular}

}

\section{Discussion}
The U.S. is alone among the countries in the industrialized world where the expected “flattening of the curve” did not take place yet. By May 31, 2020, there were 1.8 million confirmed COVID-19 cases and 99,800 deaths. 45 states were in various phases of re-opening and 5 states did not have shelter-in-place orders. By mid-June, cases had started to rise and as of June 26, there were 2.5 million confirmed cases and over 120,000 deaths. Some states that had begun to re-open parts of their economy have paused or delayed opening in the face of a surge of new cases.

Estimating the impact of mitigation strategies on cases and deaths in the U.S. is challenging particularly due to the lack of uniformity in timing, implementation, enforcement, and adherence across states. Nevertheless, early observations point to the utility of such measures, particularly shelter-in-place orders in reducing infection spread and deaths (per data from California and Washington State) \cite{washin}. Counties implementing shelter-in-place orders were associated with a 30.2\% reduction in weekly cases after 1 week, 40\% reduction after 2 weeks, and 48.6\% reduction after 3 weeks \cite{fowler2020effect} Conversely, model projections estimate a steady rise in cases and over 181,000 deaths if such mitigation strategies were to be eased and not re-enforced before October 1 \cite{washin1}.

Many researchers worldwide are currently investigating the changes in social and individual behaviors in response to the sudden yet prolonged outbreaks of COVID-19, e.g., \cite{adiga2020interplay,dascritical}. As the pandemic progresses, and until medical treatments or vaccination are available, new and diverse patterns of mobility, be they voluntary or via interventions, may emerge in each society. It is, therefore, of great importance to epidemiologists and policy-makers to understand the dynamic patterns of dependency between human mobility and COVID-19 incidence in order to precisely evaluate the impact of such measures. In this study, we have shown that such dependencies not only change over time but across locations and populations, and are possibly determined by underlying socioeconomic characteristics. Our analytical approach is particularly relevant considering the high socioeconomic costs of such measures.

We understand that our study has some limitations. We note that each step of our framework could be improved in isolation or as a pipeline, which we aim to do in our future work. We have also developed a prototype of an interactive tool to run online the steps of our analytical pipeline. It will be made publicly available shortly upon completion.

Here it is important to note the so-called ecological fallacy in inferring about individual health outcomes based on data or results that are obtained at either city or county levels. Such inference may suffer from incorrect assumptions and biases, which, however unintentional, must be avoided. Any views that might have reflected on the analysis or results of our study are those of the authors only, and not the organizations they are associated with. 

%\bibliographystyle{plain}
%\bibliography{CovidClusteringBIB}

\begin{thebibliography}{10}

\bibitem{adiga2020interplay}
Aniruddha Adiga, Lijing Wang, Adam Sadilek, Ashish Tendulkar, Srinivasan
  Venkatramanan, Anil Vullikanti, Gaurav Aggarwal, Alok Talekar, Xue Ben,
  Jiangzhuo Chen, et~al.
\newblock Interplay of global multi-scale human mobility, social distancing,
  government interventions, and {COVID-19} dynamics.
\newblock {\em medRxiv}, 2020.

\bibitem{agueh2011barycenters}
Martial Agueh and Guillaume Carlier.
\newblock Barycenters in the {W}asserstein space.
\newblock {\em SIAM Journal on Mathematical Analysis}, 43(2):904--924, 2011.

\bibitem{benamou2015iterative}
Jean-David Benamou, Guillaume Carlier, Marco Cuturi, Luca Nenna, and Gabriel
  Peyr{\'e}.
\newblock Iterative {B}regman projections for regularized transportation
  problems.
\newblock {\em SIAM Journal on Scientific Computing}, 37(2):A1111--A1138, 2015.

\bibitem{carlier2015numerical}
Guillaume Carlier, Adam Oberman, and Edouard Oudet.
\newblock Numerical methods for matching for teams and {W}asserstein
  barycenters.
\newblock {\em ESAIM: Mathematical Modelling and Numerical Analysis},
  49(6):1621--1642, 2015.

\bibitem{cuturi2014fast}
Marco Cuturi and Arnaud Doucet.
\newblock Fast computation of {W}asserstein barycenters.
\newblock {\em Journal of Machine Learning Research}, 2014.

\bibitem{dascritical}
Sarmistha Das, Pramit Ghosh, Bandana Sen, Saumyadipta Pyne, and Indranil
  Mukhopadhyay.
\newblock Critical community size for {COVID-19}: A model based approach for
  strategic lockdown policy.
\newblock {\em Statistics and Applications}, 18(1):181--196, 2020.

\bibitem{deheuvels1980non}
Paul Deheuvels.
\newblock Non parametric tests of independence.
\newblock In {\em Statistique non param{\'e}trique asymptotique}, pages
  95--107. Springer, 1980.

\bibitem{flaxman2020estimating}
Seth Flaxman, Swapnil Mishra, Axel Gandy, H~Juliette~T Unwin, Thomas~A Mellan,
  Helen Coupland, Charles Whittaker, Harrison Zhu, Tresnia Berah, Jeffrey~W
  Eaton, et~al.
\newblock Estimating the effects of non-pharmaceutical interventions on
  {COVID-19} in europe.
\newblock {\em Nature}, pages 1--8, 2020.

\bibitem{washin}
Geoffrey~A. Fowler, Heather Kelly, and Reed Albergotti.
\newblock Social distancing for coronavirus is flattening the curve,
  {C}alifornia and {W}ashington data show - the washington post.
\newblock Website, June 28 2020.

\bibitem{fowler2020effect}
James~H Fowler, Seth~J Hill, Nick Obradovich, and Remy Levin.
\newblock The effect of stay-at-home orders on {COVID-19} cases and fatalities
  in the united states.
\newblock {\em medRxiv}, 2020.

\bibitem{ghinai2020first}
Isaac Ghinai, Tristan~D McPherson, Jennifer~C Hunter, Hannah~L Kirking, Demian
  Christiansen, Kiran Joshi, Rachel Rubin, Shirley Morales-Estrada, Stephanie~R
  Black, Massimo Pacilli, et~al.
\newblock First known person-to-person transmission of severe acute respiratory
  syndrome coronavirus 2 ({SARS-CoV-2}) in the {USA}.
\newblock {\em The Lancet}, 2020.

\bibitem{washin1}
HealthData.org.
\newblock {COVID-19} projections. institute for health metrics and evaluation.
\newblock Website, Accessed June 28, 2020,.

\bibitem{house2020proclamation}
White House.
\newblock Proclamation on declaring a national emergency concerning the novel
  coronavirus disease ({COVID-19}) outbreak.
\newblock {\em White House}, 2020.

\bibitem{keogh2001derivative}
Eamonn~J Keogh and Michael~J Pazzani.
\newblock Derivative dynamic time warping.
\newblock In {\em Proceedings of the 2001 SIAM international conference on data
  mining}, pages 1--11. SIAM, 2001.

\bibitem{le2017existence}
Thibaut Le~Gouic and Jean-Michel Loubes.
\newblock Existence and consistency of {W}asserstein barycenters.
\newblock {\em Probability Theory and Related Fields}, 168(3-4):901--917, 2017.

\bibitem{NIPS2017_7062}
Scott~M Lundberg and Su-In Lee.
\newblock A unified approach to interpreting model predictions.
\newblock In I.~Guyon, U.~V. Luxburg, S.~Bengio, H.~Wallach, R.~Fergus,
  S.~Vishwanathan, and R.~Garnett, editors, {\em Advances in Neural Information
  Processing Systems 30}, pages 4765--4774. Curran Associates, Inc., 2017.

\bibitem{marti2015hcmapper}
Gautier Marti, Philippe Donnat, Frank Nielsen, and Philippe Very.
\newblock {HCMapper}: An interactive visualization tool to compare
  partition-based flat clustering extracted from pairs of dendrograms.
\newblock {\em arXiv preprint arXiv:1507.08137}, 2015.

\bibitem{inbook}
Frank Nielsen.
\newblock Hierarchical clustering.
\newblock In {\em Introduction to {HPC} with {MPI} for Data Science}, pages
  195--211. Springer, 2016.

\bibitem{pele2009fast}
Ofir Pele and Michael Werman.
\newblock Fast and robust earth mover's distances.
\newblock In {\em 2009 IEEE 12th International Conference on Computer Vision},
  pages 460--467. IEEE, 2009.

\bibitem{peyre2019computational}
Gabriel Peyr{\'e}, Marco Cuturi, et~al.
\newblock Computational optimal transport: With applications to data science.
\newblock {\em Foundations and Trends{\textregistered} in Machine Learning},
  11(5-6):355--607, 2019.

\bibitem{villani2008optimal}
C{\'e}dric Villani.
\newblock {\em Optimal transport: old and new}, volume 338.
\newblock Springer Science \& Business Media, 2008.

\bibitem{warren2020mobility}
Michael~S Warren and Samuel~W Skillman.
\newblock Mobility changes in response to {COVID-19}.
\newblock {\em arXiv preprint arXiv:2003.14228}, 2020.

\end{thebibliography}

\end{document}